%% file: ICDCS09-1.tex
\documentclass[times, 10pt,twocolumn]{article}
\usepackage{latex8}
\usepackage{times}
\usepackage{graphicx}
\usepackage{color}
\usepackage{epsfig}
\usepackage{graphicx}
\usepackage{amsmath}
\usepackage{amssymb}

\input{macros}

\begin{document}

\title{A Statistical Approach to Performance Monitoring\\
in Soft Real-Time Distributed Systems}

\author{Danny Bickson, Gidon Gershinsky, Ezra N. Hoch and Konstantin Shagin\\
IBM Haifa Research Lab, \\Mount Carmel, Haifa 31905, Israel, \\
\{dannybi,gidon,ezrah,konst\}@il.ibm.com
}

\maketitle
\thispagestyle{empty}

%
%

\begin{abstract}
Soft real-time applications require timely delivery of messages conforming to the soft real-time
constraints. Satisfying such requirements is a complex task both due to the volatile nature of distributed environments, as well as due to numerous domain-specific factors that affect message latency.
Prompt detection of the root-cause of excessive message delay allows
a distributed system to react accordingly. This may
significantly improve compliance with the required timeliness constraints.

In this work, we present a novel approach for distributed performance
monitoring of soft-real time distributed systems. We propose to employ recent distributed algorithms from the statistical signal processing and learning domains, and to utilize them in a different context of online performance monitoring and root-cause analysis, for pinpointing the reasons for violation of performance requirements.
Our approach is general and can be used for monitoring of any distributed system, and is not
limited to the soft real-time domain.

We have implemented the proposed framework in TransFab, an IBM prototype of soft real-time messaging fabric.
In addition to root-cause analysis, the framework includes facilities to resolve resource allocation problems, such
as memory and bandwidth deficiency. The experiments demonstrate that the system can
identify and resolve latency problems in a timely fashion.
\end{abstract}

\Section{Introduction}
%


The number of distributed systems with latency requirements rapidly grows.
In several domains, such as military, industrial automation, and financial markets, message latency plays a critical role. Since it is technically hard and in most cases costly to guarantee that each and every message is delivered within a predefined period of time
(hard-real time), many applications impose weaker requirements by allowing a small portion of messages to exceed their deadline (soft real-time).
Still, as a consequence of application complexity and the volatile nature of the distributed environment, even compliance with these weaker
constraints is a challenging task. Unexpected activity bursts,  message loss due to unreliable communication medium, network buffer overflow,
network congestion, resource sharing and many other unpredictable factors may result in significant increase in the end-to-end message delay.

It is highly desirable that a distributed system adapts to the changing conditions and thus avoids violations of the latency constraints. A crucial step towards achieving this is enabling the system to identify the root-cause whenever there is a degradation in performance. This, by itself, is a non-trivial problem, because in this context a symptom may be easily mistaken for the real cause or misinterpreted. For example, packet drop resulting from
buffer space deficiency on the receiver side may be attributed to packet loss due to network congestion.

Distributed systems with latency constraints often employ resource reservation to ensure that the more critical components are served more promptly. The reserved resources are commonly the memory space, bandwidth and CPU share. In many cases, readjustment of the resource quotas can alleviate the timeliness issues. For instance, if a certain component rapidly generates messages, its may exceed its transmission bandwidth limit
and hence may have to queue messages, rather than transmitting them immediately. Consequently, the delayed messages may miss their delivery deadline. This may be avoided by temporary increasing the component's bandwidth, if possible.

The ideas above lead us to devise a framework that monitors distributed system performance, determines the root-cause of the increased delay, and takes corrective actions in order to avoid violation of the timeliness constraints.  We propose a monitoring framework which employs a distributed root-cause analysis. A significant advantage of the statistical approach is that, in contrast to the expert knowledge methods, it is independent of the system characteristics such as operating system, transport protocol and network structure. Moreover, it requires a minimal domain-specific knowledge to accurately determine the root-cause.

Our primary design goals were the following:
\begin{itemize}
\item System operation should be distributed, without a centralized computing node.
\item The system should adapt to network changes as quickly as possible.
\item The system should not rely on software implementation, OS and networking details (``black-box'' approach).
\end{itemize}

In the current work, we make the novel contribution of borrowing recent algorithms from the field of statistical signal processing~\cite{Allerton08-2,ISIT1} to be employed in a different context of a distributed monitoring framework. By utilizing those algorithms we are able to efficiently and distributively characterize the behavior of the varying network conditions as a stochastic process, and to perform root-cause analysis for detecting the parameters which cause an increased latency.

The framework works as follows. Each node monitors a large number of various local operating system and application  parameters. If a degraded performance is observed anywhere in the network, the nodes jointly characterize the performance by regarding it as a linear stochastic process, using statistical signal processing tools. Subsequently,
a joint root-cause analysis computation is performed to identify the parameters which affect performance.
Once the reasons for degradation are known, a corrective action is taken (whenever possible), by adjusting the resource quota of one or more nodes.

The root-cause analysis technique is general and can be applied in many other distributed systems,
and it is not limited to the soft real-time domain. The main performance measures are tunable and can be
set, for example, to CPU consumption, bandwidth utilization etc. One of the appealing properties of our monitoring is that it can be used for debugging as well -- detecting anomalous software behavior
like bugs and deadlocks. It can be further used for load balancing, minimization
of deployed resource, hot-spot detection etc.

We have implemented the proposed framework in TransFab, a prototype of soft
real-time messaging transport fabric, developed in IBM Research Lab.
We have tested our framework in various settings and on different topologies.
The experiments show that the proposed scheme accurately identifies the reasons for performance
degradation in non-trivial scenarios. Overall, the protocol is a light-weight protocol. The message
overhead of a single root-cause analysis computation amounts to only several kilobytes per communicating node, which is negligible in most contemporary networks. We have further observed only a minor increase in CPU consumption and
memory. 

Our technique can scale up to large domains, in a hierarchical manner, where each sub domain performs monitoring locally, filters out the relevant parameters which affect performance, and then the algorithm is run again between the different domains.

This paper is organized as follows. Section~\ref{sec:related} describes related previous work.
Section~\ref{sec:background} outlines the mathematical background required for understanding our construction.
Section~\ref{sec:const} presents our framework. Experimental results of a real LAN deployment
are discussed in Section~\ref{sec:exp}. We conclude in Section~\ref{sec:conc}.

\Section{Related work}
\label{sec:related}
Recently, there has been a lot of research targeted on
monitoring and allocation of resources
in communication networks utilizing techniques from
statistics, learning and data mining domains (see for
example \cite{Gerry1,Gerry2,ICDCS08-1,ICDCS08-2,ICDCS08-3,
ICDCS08-4}). One possible approach for having strict performance guarantees of a distributed software application is to use over provisioning, where the required host and network resources are allocated ahead~\cite{ICDCS08-3}, avoiding
cases of resource congestion. In contrary, in the current work, we assume a dynamic model, where software behavior and resource requirements are not known ahead.

Other works use a centralized computation for computing the best allocation of resources~\cite{Gerry1,ICDCS08-2},
while we assume a distributed computing model. In our model there is no central server where all the information
is shipped and processed in. Rish \etal~\cite{Gerry2} optimizes topology construction for
optimizing download speed of a Peer-to-Peer network. In the current work, we assume there is some given topology of the communication flows between the participating nodes as designed by the application builder, and perform the monitoring on top of the given topology. Our monitoring framework can be applied in any given topology, including graphs with cycles.


Resource Bundles~\cite{ICDCS08-4} is an example of a successful approach for finding and clustering similar available resources over the WAN, and is closely related to this work. In the ResourceBundles system, resources are captured daily to form a resource utilization histogram. Available resources are aggregated for providing users smart selection of groups of resources.
Not surprisingly, it is shown that adding historical information about resource consumptions improves significantly selection of resources. In the current paper, our focus is on real time resource allocation: we capture resource usage in a time window of seconds (rather than daily). Furthermore, we use a characterization of the process as a linearly noisy stochastic process. This allows greater flexibility in describing the process behavior over time, by characterizing joint covariance of measured parameters (relative to histograms used by ResourceBundles). Furthermore, our framework is also capable of performing root-cause analysis of parameters which affect system performance.

\Section{Mathematical Background}
The following Section briefly overviews the mathematical background needed for describing
the algorithms deployed. Section~\ref{sec:const} explains how those algorithms are used in the monitoring
context.
\label{sec:background}
\SubSection{The Kalman filter algorithm}
\label{sec:kalman}
The Kalman Filter is an efficient iterative
algorithm that estimates the state of a discrete-time controlled
process $x \in R^n$ that is governed by the linear stochastic
difference equation \BE
x_k = Ax_{k-1} + w_{k-1}, \EE \
with a measurement $z \in R^m$ that is $ z_k = Hx_k + v_k.$ The
random variables $w_k$ and $v_k$ that represent the process and
measurement AWGN noise (respectively). $p(w) \sim \mathcal{N}(0,
Q), p(v) \sim \mathcal{N}(0, R)$.
The discrete Kalman filter update equations are given
by~\cite{Kalman}:

The prediction step:
\begin{subequations}
\begin{eqnarray}
 \hat{x}^-_k &=& A\hat{x}_{k-1}, \label{hat_x_minus_k}\\
 P_k^- &=& AP_{k-1}A^T + Q. \label{p_k_minus}
\end{eqnarray}
\end{subequations}

The measurement step:
\begin{subequations}
\begin{eqnarray}
 K_k &=& P_k^-H^T(HP_k^-H^T + R)^{-1}, \label{kalman_gain} \\
\hat{x}_k &=& \hat{x}^-_k + K_k(z_k - H\hat{x}^-_k),
\label{hat_x_k} \\
 P_k &=& (I-K_kH)P_k^-. \label{p_k}
\end{eqnarray}
\end{subequations}
where $I$ is the identity matrix.

The algorithm operates in rounds. In round $k$ the estimates
$K_k,\hat{x}_k,P_k$ are computed, incorporating the (noisy)
measurement $z_k$ obtained in this round. The output of the
algorithm are the mean vector $\hat{x}_k$ and the covariance
matrix $P_k$.

\SubSection{Generalized least squares (GLS) method}
\label{sec:GLS}
Given an observation matrix $A$ of size $n \times k$, and a target vector $b$ of size $1 \times n$, the linear
regression computes a vector $x$ which is the least squares solution to the quadratic cost function
 \[ \min_x||Ax - b||^2_2\;. \] The algebraic solution is $x = (A^TA)^{-1}A^T b$. $x$ can be referred as the hidden weight parameters, which given the observation matrix $A$, explains the target vector $b$.

The linear regression method has an underlying assumption that the measured parameters are not correlated.
However, as shown in the experimental results in Section~\ref{sec:exp}, the measured parameters are highly correlated.
For example, on a certain queue the number of get/put operations in each given second are correlated. In this case,
it is better to use the generalized least squares (GLS) method. In this method, we minimize the quadratic cost function
\BE \min_x (Ax-b)^TP^{-1}(Ax-b)\;, \label{ls} \EE where $P$ is the inverse covariance matrix of the observed data. In this case, the optimal
solution is \BE x = (A^TP^{-1}A)^{-1}A^TP^{-1}b \;. \label{blue} \EE which is the best linear unbiased estimator (BLUE).


\SubSection{Efficient distributed computation via the Gaussian Belief Propagation algorithm}
\label{sec:gabp}
Recent work~\cite{Allerton08-2} shows how to compute distributively and efficiently the Kalman filter over a communication network.

Other recent work~\cite{Allerton,ISIT1,ISIT2} show that the GLS method computation~(Eq. \ref{blue}) can be computed efficiently and distributively the GaBP algorithm as well.

The Gaussian Belief Propagation (GaBP) algorithm is an efficient iterative algorithm for solving a system
of linear equations of the type $Ax=b$~\cite{ISIT1}. The input to the algorithm is the matrix $A$ and the vector $b$, the output is the vector $x = A^{-1}b$. The algorithm is a distributed algorithm, which means that each node gets a part of the matrix $A$ and the vector $b$
as input, and outputs a part of the vector $x$ as output. The algorithm may not converge, but in case it converges
it is known to converge to the correct solution.

Because of the short space, we do not reproduce here any of the previous results. The interested reader is referred
to~\cite{Allerton08-2,ISIT1,ISIT2,Allerton} for complete description of the algorithms deployed.


\section{Our proposed construction}
\label{sec:const}
\begin{figure}[ht!]
\begin{center}
  \includegraphics[scale=0.66]{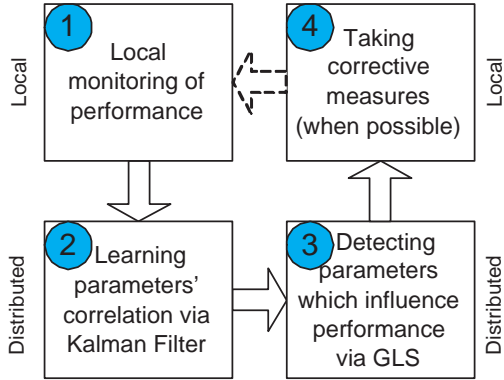}\\
  \caption{Schematic operation of the proposed monitoring framework.}\label{fig:schema}
\end{center}
\end{figure}

Our monitoring framework is composed of four stages, as depicted in Figure~\ref{fig:schema}. 
The first stage is the data collection stage. In this stage the node locally monitor their measurable performance
parameters and record the relevant parameters in a local data structure. The data collection is done in the background every
configured time frame $\Delta t$, and has minimal effect on performance. In case of a normal operation, all messages
arrive before their soft real-time deadlines. Thus, there is no need to continue and compute the next stages.
Whenever one of the nodes detects some deterioration in performance (e.g., a message is almost late), it notifies the other nodes that it wishes to compute the second stage.

The second stage performs the Kalman filter computation
distributively. The input to the second stage are the
local data parameters collected in the first stage,
and its output is the mean and joint covariance matrix
which characterize correlation between the different parameters (possibly collected on different machines). The underlying algorithm
used for computing the Kalman filter updates is the GaBP algorithm (described in Section~\ref{sec:gabp}).
The output of the second stage can be also used for reporting performance to the application. For example, we
are able to measure the mean and variance of the effective bandwidth.

The third stage computes the GLS method (explained in Section~\ref{sec:GLS}) for performing regression. The target for the regression can be chosen on the fly. In our experiments we where mainly interested in the total message latency as our most important performance measure. The input to the third stage is the parameter data collected at the first stage, and the covariance matrix computed in the second stage. The output of the third stage is a weight parameter vector.
The weight parameter vector has an intuitive meaning of providing a linear model for the data collected. The computed linear model allows us to identify which parameters influence performance the most (parameters with the highest absolute weights). Additional benefit, is that using the computed weights we are able to compute predictions for the node
behavior. For example, how an increase of 10MB of buffer memory will affect the total latency experienced.

Finally, the fourth stage uses the output of the third stage for taking corrective measures. For example, if the
main reason of increased latency is related to insufficient memory, the relevant node may request additional memory resources from the operating system. The fourth stage is done locally and is optional, depending on the type of application and the availability of resources.

Below, we give further details regarding the implementation and computational aspects of the different stages.

\SubSection{Stage I: local data collection}
In this stage, participating nodes locally monitor their performance every $\Delta t$ seconds. Node record performance parameters, such as memory and CPU consumption, bandwidth utilization and other relevant parameters. Based on the monitored software, information about internal data structures
like files, sockets, threads, available buffers etc. is also monitored. The monitored parameters
are stored locally, in an internal data structure representing the matrix $A$, of size $n \times k$, where $n$ is the history
size, and $k$ is the number of measured parameters. Note, that at this stage, the monitoring framework is oblivious to the meaning of the monitored parameters, regarding all monitored parameters equally as linear stochastic noisy processes.

\SubSection{Stage II: Kalman filter}
The second stage is performed distributively over the network, where participating nodes compute the Kalman filter
algorithm (outlined in Section~\ref{sec:kalman}). The input to the computation is the matrix $A$ recorded in the data collection stage, and the assumed levels of noise $Q$ and $R$. The output of this computation is the mean vector $\hat{x}$ and the joint covariance matrix $P$ (Eq.~\ref{hat_x_k},~\ref{p_k}). The joint covariance matrix characterizes correlation between measured parameters, possibly spanning different nodes.

We utilize recent results from the field of statistical signal processing for computing the the Kalman filter
using the GaBP iterative algorithm (explained in Section~\ref{sec:gabp}.) The benefit of using this recent efficient distributed iterative algorithm is in faster convergence (reduced number of iterations) relative to classical
linear algebra iterative methods. This in turn, allows the monitoring framework to adapt promptly to changes in the network.

The output of the Kalman filter algorithm  $\hat{x}$ is computed in each node locally. Each computing node has the part of the output which is the mean
value of its own parameters. For reducing computation cost, we do not compute the full matrix $P$, but the rows of $P$ which represent significant performance parameters selected ahead. 

\SubSection{Stage III: GLS Regression}
The third stage is performed distributively over the network as well, for computing the GLS regression (Eq.~\ref{blue}). The input to this stage is the joint covariance matrix $P$ computed in the second stage, the
recorded parameters matrix $A$, and the performance target $b$. The output of the GLS computation is a weight
vector $x$ which assigns weights to all of the measured parameters. By selecting the parameters
with the highest absolute magnitude from the vector $x$, we identify which of the recorded parameters
significantly influence the performance target. The results of this computation is received locally, which means
that each node computes the weights of its own parameters. Additionally, the nodes compute distributively
the top ten maximal values. The GLS method is computed again using the GaBP algorithm (Section~\ref{sec:gabp}).
The main benefit of using the GaBP algorithm for both tasks (kalman filter and GLS method computation) is that we need to implement and test the algorithm only once.

\SubSection{Stage IV: taking corrective measures}
\label{prediction}
Whenever a node detects that a local parameter computed in stage III, is highly correlated to the target
performance measure, it may try to take corrective measures. This step is optional and depends on the application and/or the operating system
support. Example of local system resources are CPU quota, thread priority, memory
allocation and bandwidth allocation. Note that resources may be either increased or decreased based on the regression results.

For implementing this stage, a mapping between the measured parameters and the relevant resource needs to be
defined by the system designer. For example, {\em TRANSMITTER\_PROCESS\_VSIZE }, the process virtual memory size is related
to memory allocated to the process by the operating system. Our monitoring framework (stages I - III) is not aware to the semantic meaning of this parameter. For taking corrective measures, the mapping between parameters
and resource is essential and requires domain specific knowledge. Getting back the virtual memory example above,
the mapping links {\em TRANSMITTER\_PROCESS\_VSIZE} to the memory quota of the transmitter process. Whenever 
this parameter is selected to by the linear regression done in stage III as a parameter which significantly
affects performance, a request to the operating system to increase the virtual memory quota is performed. 

A natural question is how much to increase / decrease a certain resource quota. Here, the results of Stage III are useful. The regression assign weights to examined system parameters to explain the performance target in the linear model. More formally, $Ax \thickapprox b$ where $x$ is the weight vector, $A$ are the recorded parameters and $b$ is the performance target. Now, assume $x_i$ is the most significant parameter selected by the regression, representing resource $i$. It is possible to increase $x_i$ by let's say 20\%, $\hat{x} = x + 0.2*x_i$ and examine the result of the increase on the predicted performance, by using the equation $\hat{b} = A\hat{x}$.

Specifically, in the soft real-time systems we focus on, we can examine the effect of an increase of 10\% in transmitter memory by computing the predicted effect on total message latency. In the current work we
mainly experimented with memory predictions, where increase was limited to up to 10\%. We have found the linear
model quite accurate under those settings, whenever the memory was the actual performance bottleneck. An area for future work is to investigate the applicability of predictions computed by the linear model on broader settings and applied to other resources.

\Section{Experimental Results}
\label{sec:exp}
The TransFab messaging fabric is a high-performance soft real-time messaging middleware.
It runs on top of the networking infrastructure and implements a set of publish/subscribe and point-to-point services
with explicitly enforced limits on times for end-to-end data delivery. 

We have incorporated our monitoring framework as a part of the TransFab overlay in Java.
In our experiments, the TransFab node recorded 190 parameters which characterize the current performance. Among them, memory usage, process information (obtained from the {\tt $\backslash$proc} file system in Linux), current bandwidth, number of incoming/outgoing messages, state of internal data structures like queues and buffer pools, number of get/put operations on them, etc.

We have utilized the unreliable UDP transport whose timeliness properties are more predictable then those of TCP. TransFab incorporates reliability mechanisms that guaranties in-order delivery of messages.
A transmitter discards a message only after all receivers have acknowledged the receipt of the message. When a receiver detects a missing message, it requests its retransmission by sending a negative acknowledgement to the transmitter.

\SubSection{Two nodes experiment}
For testing our distributed monitoring framework, we have performed the following small experiment.
In this experiment, our main performance measure is the total packet latency.
A transmitter and receiver TransFab nodes run on two idle Pentium IV dual core AMD Opteron 2.6Ghz Linux machines on the same LAN. The transmitter was configured to send 10,000 messages of size 8Kb in a second. Memory allocation of both nodes is 100Mb.
The experiment run for 500 seconds, where history size $n$ was set to 100 seconds. During the experiment, stage I (data collection) of the monitoring was performed every $\Delta t = 1$ second. At time 250 seconds, the Kalman filter
algorithm was computed distributively by the nodes. The goal of this
experiment is to show that by performing the Kalman filter computation (stage II) using information collected
from two nodes, we are able to identify which of the collected parameters influence the total packet latency.
Furthermore, we are able to gain insights about system performance, which could not be computed by using only local
information.

\begin{figure}[ht!]
\hspace{-1cm}
  \includegraphics[scale=0.3]{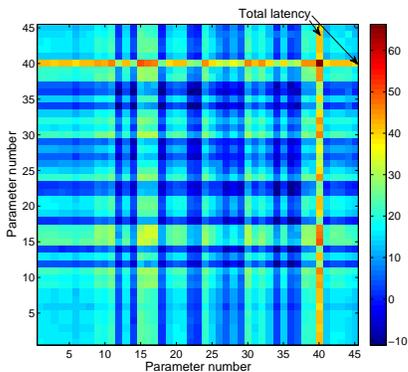}\\
  \caption{Joint Covariance matrix computed distributively using two TransFab nodes. Column (and row) 40 captures the dependency of total packet latency in various measured parameters. Warm colors (yellow to red) presents medium to high correlation of measured parameters with the total latency.}\label{covariance2}
\end{figure}

For saving bandwidth, nodes locally filter out constant parameters out of the matrix $A$. Thus, the input
to the Kalman filter algorithm was reduced to 45 significant parameters. Figure \ref{covariance2} presents a joint covariance matrix calculated by the Kalman filter algorithm (computed in the second stage) using this
typical run. Column (and row) 40 represent the total packet latency measured by the receiver. The covariance matrix includes parameters captured by the transmitter (columns 1-23) and parameters recorded by the receiver (columns 24-45).

As clearly seen in Figure \ref{covariance2}, the total
latency of packets, even in a small setup of only two nodes,
on two idle machines, is strongly correlated with dozens of
parameters. Furthermore, the total latency depends on
parameters from both the sender and the receiver.
The covariance matrix plots the dependence of pairs of
parameters, where we are mainly interested in understanding
the reasons for message delay. 

In practice, we have deployed
the following optimization: we do not compute the full covariance matrix, but only the rows which represent correlation with the target parameters (in this example only row 40). 

\begin{figure}[ht!]
\begin{center}
  \includegraphics[scale=0.27]{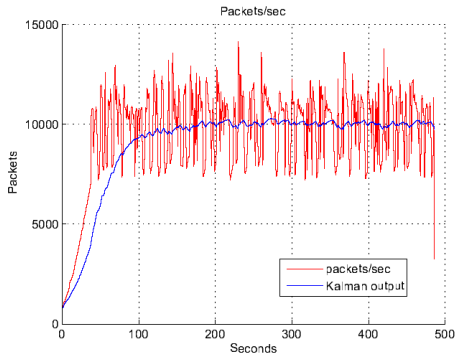}\\
  \caption{Kalman filter smoothing of packets per second parameter.}\label{fig:packets_per_sec_kalman}
\end{center}
\end{figure}

Figure~\ref{fig:packets_per_sec_kalman} depicts the additional Kalman filter output, which is the mean vector $x_k$.
In this example, the mean packets per second parameter of the same two nodes experiment.
The assumed error levels $Q,R$ define the level of smoothing. In our experiments we took $Q,R$ to be diagonal matrices with error level $\sigma^2 = 0.01$. The mean value and computed variance provide the nodes with additional
information about performance, which could be used for monitoring and debugging.

\begin{figure}[ht!]
\begin{center}
 \includegraphics[scale=0.66]{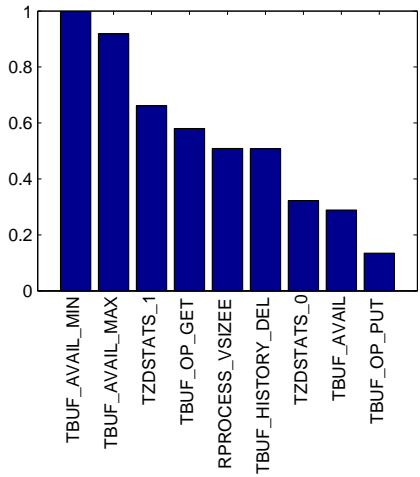}\\
  \caption{Linear regression results of a transmitter with low memory. Out of the first 5 top reasons for latency increase are 3 related to the transmitter memory buffers.}
\label{fig:lin_reg}
\end{center}
\end{figure}

\begin{figure}[ht!]
\begin{center}
 \includegraphics[scale=0.66]{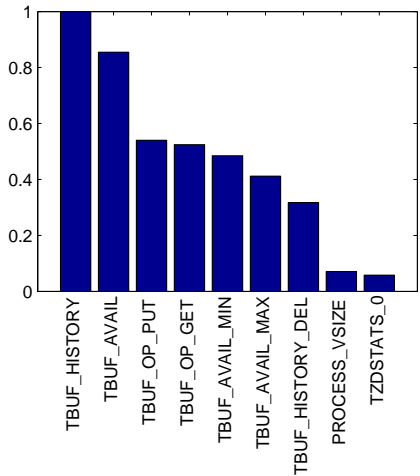}\\
 \caption{Improved regressions results for a transmitter with low memory, using the GLS method. Seven out of nine top reasons for latency increase are related to the transmitter memory buffers.}
\label{fig:gen_lin_reg}
\end{center}
\end{figure}

We have repeated the previous experiment, but this time at time 150 seconds, the transmitting machine memory was reduced to 2.4Mb for outgoing message buffers. At time 155, the receiving nodes detected locally degradation in performance, and compute stages II (Kalman filter) and III (linear regression) where the history parameter was set to $n=100$ seconds, for finding the parameters which affect the total packets latency. Figure~\ref{fig:lin_reg} presents the output of the
distributed linear regression preformed by the two TransFab nodes. Clearly, the transmitter low buffers are the main (3 out of the top 5) reasons behind the increased latency. We deployed the GLS method (shown in Figure~\ref{fig:gen_lin_reg}) to improve the quality of ranking. This time, all seven out of the top 9 parameters are related to transmitter buffers.

\SubSection{Larger experiments}
We have performed the following experiment which demonstrates the applicability of our monitoring framework.
The goal of this experiment was to test, given a randomly chosen faulty node, weather the monitoring framework
is able to correctly identify the faulty node, and the type of the fault.

\begin{table}
  \centering
\begin{tabular}{|l|l|}
  \hline
  Error & Description \\ \hline
  A & No error \\
  B & Low CPU receiver \\
  C & Low CPU transmitter \\
  D & Channel loss\\
  E & Low memory receiver \\
  F & Low memory transmitter \\
   \hline
\end{tabular}
  \caption{Possible host and network faults selected randomly on runtime.}\label{tab:kfaults}
\end{table}

At runtime, we have randomly created an overlay tree topology of ten nodes. A single transmitter (the root note), transmits a flow of 3,000 packets of size 8K per second to its children, with a rate of 24Mbit for each flow. The nodes have no global knowledge of the network topology. Specifically, the transmitter is not aware which of its direct
children are forwarding nodes and which nodes are tree leaves. Next, we have selected at random one of faults listed in Table~\ref{tab:kfaults} to be assigned to one of the tree nodes.

We have run the experiment for 500 seconds, after which the nodes jointly compute the regression results, where the target was the transmitter latency. (Transmitter latency is defined as the total time messages wait in transmitter buffers from their submission by the application until they are actually transmitted over the wire.)

\begin{figure}[ht!]
\begin{center}
  \includegraphics[scale=0.45]{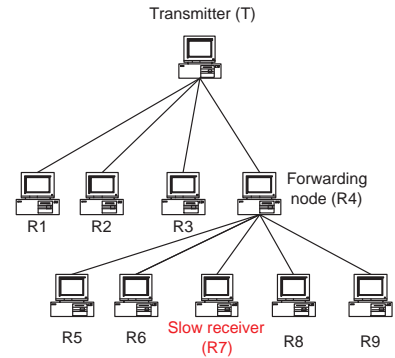}\\
  \caption{A tree topology of ten machines. The topology, forwarding node and type of error (in this case low memory receiver) are randomly selected at runtime. Regression is performed jointly on all ten nodes, where the target is the transmitter latency.}\label{fig:topology}
\end{center}
\end{figure}

\begin{figure}[ht!]
\begin{center}
  \includegraphics[scale=0.45]{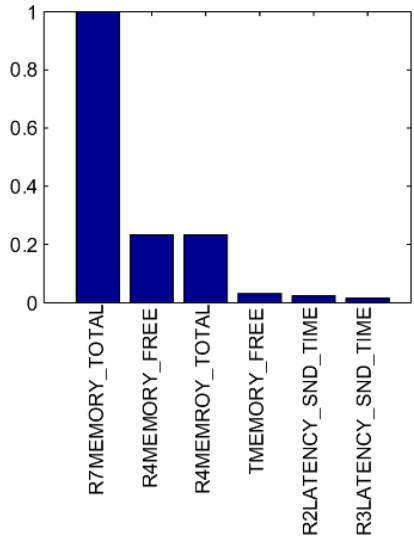}\\
  \caption{Regression results for the tree topology with ten nodes. The low memory receiver (R7) memory is detected to be the first cause of transmitter latency. The forwarding node's memory (R4) is the second cause for transmitter latency. Finally, transmitter (T) memory is detected as the fourth cause which affects transmitter latency. }\label{fig:9nodes_reg}
\end{center}
\end{figure}

%
This experiment was repeated multiple times, each time with a different topology, a different faulty node and a different fault was selected. An example topology generated at random is shown in Figure~\ref{fig:topology}. The faulty node was assigned fault $E$ - a low memory receiver. The matching regression results are presented in Figure~\ref{fig:9nodes_reg}. The results indicate, that the memory of the low memory node is identified as the first cause which affects transmitter latency. The forwarding node's memory is identified as the second and third causes of the transmitter latency. Finally, the transmitter memory is identified as the fourth cause of transmitter latency.
Besides of detecting the faulty node, the critical path between the transmitter and the low memory receiver
is identified correctly as the congested path.

\begin{figure}[ht!]
\hspace{-0.8cm}
  \includegraphics[scale=0.3]{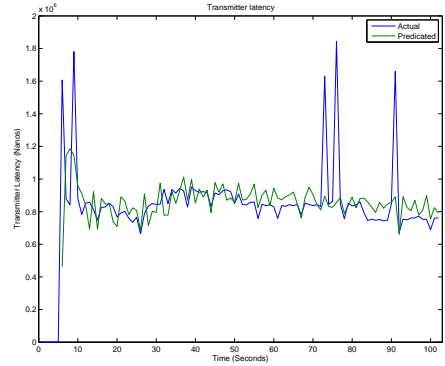}\\
  \caption{Regression quality for the ten node experiment. The blue line represents actual transmitter latency, while the green line represents the predicted latency using the linear model.}\label{fig:sender_latency}
\end{figure}

Figure~\ref{fig:sender_latency} depicts the quality of the linear regression for the same experiment, comparing the actual transmitter latency with the predicted latency by the linear model. More formally, we first compute $x$ using Eq.~\ref{blue} and then plot $Ax$ vs. $b$. The desired result is that the actual latency and predicted latency
computed by the linear model would be as close as possible to each other. However, using real world data it is hard
to get perfect predictions, probably because the linear model is a simplification of the real world.
This figure shows that the overall fit between the predicted and actual latency is quite good, except for some spikes
in the actual transmitter latency which where not predicted. Note that prediction quality is closely related to the discussion in Section~\ref{prediction} regarding resource allocation in stage IV. 


We have repeated the ten nodes experiment multiple times, each time selecting a different topology at random
and a different fault. Table~\ref{tab:results} summarizes our findings.
\begin{table*}[ht!]
  \centering
\begin{tabular}{|l|c|c|c|}
  \hline
  Error & Faulty node identified? & Reason for fault identified? & Reason for fault identified\\
  & & & using domain knowledge? \\ \hline
  A (no error) & - & - &\\
  B (receiver CPU) & V & B,D undistinguishable & V\\
  C (transmitter CPU) & V & C,E undistinguishable & V\\
  D (channel loss) & receiver &  B,D undistinguishable & V\\
  E (receiver memory) & V & V & \\
  F (transmitter memory) & V & V & \\
   \hline
\end{tabular}
  \caption{Summary of results for the randomly constructed ten nodes overlays.}\label{tab:results}
\end{table*}
In all cases we where able to correctly identify the faulty node, except of the channel loss fault D.
The channel loss case simulates a lossy channel by dropping packets uniformly at random with
a probability of 5\%. In the case of channel loss, the identified node was the receiver.

Additional question we ask, is weather the linear regression is able to identify the type of fault.
The easiest faults to detect where low memory faults (E+F), where either the transmitter or receiver had low memory. In those cases both the node and the fault reasons where identified correctly. The case of normal behavior (fault A) we identified receivers latency to have highest correlation to transmitter latency, which is normal.

The loss situation (fault D) where much harder to distinguish from a low CPU receiver (fault B). The reason
is that it is much harder to enforce flow control using IP multicast, since different nodes have different capabilities. In the case of fault B, the slow receiver is swamped with packets in speed higher than its processing capabilities, so packet
loss is incurred at the operating system socket buffers level. From the other hand, in fault D, random packets
are thrown in a situation which is similar to fault B. Table~\ref{tab:results} summarizes the groups of faults that where not distinguishable using the linear regression.

Since we where not able to distinguish between channel loss and socket buffer overflow, we experimented
with bursty loss model (instead of random loss). Under the bursty loss model, we have thrown a sequence of 100 packets, with a probability of \%1 chosen randomly in uniform. In this case it was much easier to distinguish between faults
B and D by comparing the pattern of negative acknowledgements.

We conclude that the linear regression is a very effective tool for identifying the faulty nodes as well
as the critical congested path in a data dissemination overlay. However, in some cases it is harder to detect
the reason for performance degradation without deploying additional tools. We propose to use domain specific
knowledge using the data locally collected at the faulty node in those cases. One option is to record normal
behavior of the node locally (computed in stage II), so a faulty node can compare mean and variance for identifying
locally anomalous behavior. Another option is to add expert knowledge, for example detecting messages which arrive
out of sequence in large gaps for identifying bursty loss cases.



\SubSection{Protocol overhead}
Regarding the protocol overhead, stages I and IV are performed locally with minimal computational effort, requiring no
network bandwidth. Stage II and III are performed across the network. The GaBP algorithm typically converges in five iterations, where in each iteration each node sends around 2.5Kb of data to each of its neighbors.  The GLS method
requires only a single execution of the GaBP algorithm, where the Kalman filter requires several executions (depends
on the required accuracy). We typically set the number of Kalman filter iterations to 10. The memory and CPU
requirements where found to be negligible. Speed of the linear regression computation was around 0.1 second, while
the Kalman filter step computation took up to 1 second. The computation are done by a separate thread, the avoid
service interruption.

\Section{Conclusion and future work}
\label{sec:conc}
In this work we have proposed an efficient monitoring framework to be used in a distributed system.
Using our approach, we regard nodes' performance parameters as a stochastic process and use statistical signal processing tools for characterizing the process behavior. Furthermore, we are able to perform a root-cause analysis across the network for identifying anomalous behaviors and parameters which affect performance.

Using a prototype software implemented in Java which ran distributively on up to ten computing nodes, we where able
to demonstrate that we are able to correctly identify faulty nodes in a randomly generated data dissemination overlay, and in most cases identify the reasons for performance degradation. However, in some difficult cases it was harder to
distinguish between several possible faults. In this case, we propose to use domain specific knowledge locally
for identifying correctly the type of fault.

As for future work, we plan to extend the local correction done at stage IV for creating a self-healing network.
Currently we focus on low memory situations where the faulty node increases its memory allocated.
We further plan to explore changes in local resource quotas like bandwidth and process priority. Another interesting
extension is to deploy hierarchy in larger networks where the monitoring is done locally in each domain, and performance results are aggregated using domain hierarchy.

The GaBP algorithm is an iterative algorithm which may not converge. We are now working on a variant of the algorithm
which always converges to the correct answer. This variant will be reported in a near future work.
\bibliographystyle{latex8}
\bibliography{ICDCS09-1}

\end{document}

%% file: macros.tex
\newcommand{\ignore}[1]{}
\newcommand{\comment}[1]  {}

\newcommand\etal{{\textsl{et al.\,}}}
\def\BE{\begin{equation}}
\def\EE{\end{equation}}
\def\BEA{\begin{eqnarray}}
\def\EEA{\end{eqnarray}}
\newcommand{\cut}[1]{{}}